\documentclass[12pt]{article}

  \usepackage{graphicx}
  \graphicspath{{../Figs/}}
  \DeclareGraphicsExtensions{.pdf,.png, .jpg}
  \usepackage[margin=1in]{geometry}

  \usepackage{bm}
  \newcommand{\beq}{\begin{equation}}
  \newcommand{\eeq}{\end{equation}}

  \usepackage[nomargin,inline,final]{fixme}
  \fxusetheme{color}
  \usepackage{hyperref}
  \hypersetup{
    unicode=false,
    pdftoolbar=true,
    pdfmenubar=true,
    pdffitwindow=false,
    pdfstartview={FitH},
    pdfcreator={pdflatex},
    pdfproducer={Latex with hyperref},
    pdfnewwindow=true,
    colorlinks=true,
    linkcolor=red,
    citecolor=green,
    filecolor=magenta,
    urlcolor=blue}

\usepackage{textcomp}
\begin{document}


  \title{Microring resonator-coupled photoluminescence\\ from silicon W~centers}

  \author{A.~N.~Tait$^{1,*}$,
          S.~M.~Buckley$^{1}$,
          J.~Chiles$^{1}$,
          A.~N.~McCaughan$^{1}$,\\
          S.~Olson$^{2}$,
          S.~Papa~Rao$^{2,3}$,
          S.~Nam$^{1}$,
          R.~P.~Mirin$^{1}$,
          J.~M.~Shainline$^{1}$
          }
  \date{}
  \maketitle

  \
  {$^{1}$ Applied Physics Division, National Institute of Standards and Technology,\\ Boulder, CO 80305, USA}

  \
  {$^{2}$ NY CREATES, Albany, NY 12203, USA}

  \
  {$^{3}$ SUNY Polytechnic Institute, Albany, NY 12203, USA}

  \

  \vspace{10pt}
  {$^*$alexander.tait@nist.gov}

\begin{abstract}
  Defect centers are promising candidates for waveguide-integrated silicon light sources.
  We demonstrate microresonator- and waveguide-coupled photoluminescence from silicon W~centers. Microphotoluminescence measurements indicate wavelengths on-resonance with resonator modes are preferentially coupled to an adjacent waveguide.
  Quality factors of at least 5,300 are measured, and free spectral ranges closely match expectation. The W~center phonon sideband can be used as a spectral diagnostic for a broader range waveguide-based devices on cryogenic silicon photonic platforms.
\end{abstract}

%
%
%
%
%


\section{Introduction}
  The rapid growth of the silicon photonics industry promises to bring manufacturing economies that are unprecedented for any non-electronic technology. In addition to potentials for large-volume production, silicon photonics opens possibilities for large-scale photonic processing architectures that could not be conceived of in fiber or III-V platforms~\cite{Perez:18,Shainline:17,Tait:17}. In all photonic systems, light sources are required. Silicon does not readily emit light at room temperature due to its indirect bandgap.
  Therefore, most efforts in silicon photonics use external light sources coupled on-chip with fiber. The use of external light sources presents critical burdens of fiber packaging and fiber-to-chip insertion loss.

  Substantial research has been dedicated to developing integrated light sources for silicon photonics~\cite{Zhou:15}. Each approach has advantages and drawbacks. These include rare earth element doping (low brightness), wafer bonding of III-V quantum wells~\cite{Roelkens:10} (non-monolithic integration steps), epitaxial growth of III-V quantum dots~\cite{Liao:18} (specialized epitaxy steps), and bandgap engineering of germanium~\cite{Bao:17} (low-yield strain engineering). All of these approaches with the exception of germanium involve the introduction of materials that are incompatible with typical silicon foundries.

\begin{figure}[tb]
    \includegraphics[width=.8\linewidth]{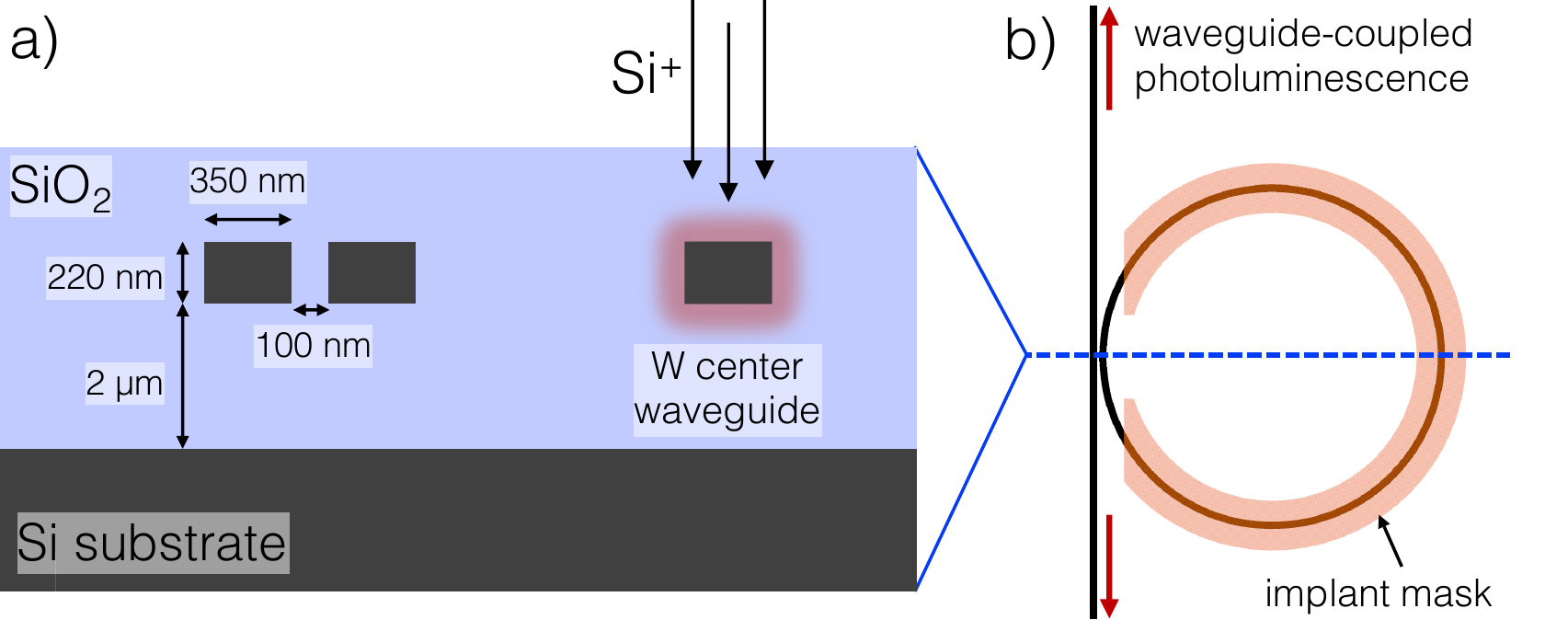}
    \caption{Illustration of platform and devices. a) Waveguide cross-sections. All waveguides are 220~nm thick and 350~nm wide such that they are single mode at 1218~nm. W-center emitters are created by implanting a waveguide with silicon ions and annealing. Evanescent couplers between microrings and the bus waveguide have 100~nm gaps in this work. b) Top view of microring resonator coupled to a waveguide. Black is silicon; red is the pattern of ion implants; blue, dashed line is the cross-section slice. Ions that miss the silicon have no effect. Implant regions are widened to relax alignment tolerance.}
    \label{fig:xsection}
  \end{figure}

  Silicon emitters can be realized by creating specific defects in the silicon crystal.
  Emissive defect centers in silicon have been studied for seven decades~\cite{Davies:89thesis}, reviewed in~\cite{Shainline:07}, but they have received little attention as sources for silicon photonic circuits. Defect centers are relatively simple to fabricate because they involve locally modifying the native crystal as opposed to introducing new materials and/or structures.
  In this work, we study W~centers in silicon-on-insulator (SOI) photonic integrated circuits, illustrated in Fig.~\ref{fig:xsection}. W~centers are trigonal defects that emit strongly at their zero phonon line at 1218~nm. They are created through Si$^+$ ion bombardment followed by an anneal.

  Like all silicon light sources, W~centers have a drawback, namely that they only operate at temperatures below 45~K.
  For applications where cryogenic operation is acceptable or required, this is not necessarily a drawback. Monolithic silicon light sources may provide a compact, inexpensive, scalable solution for superconducting optoelectronic neural networks~\cite{Shainline:17} and optical quantum information systems.
  In many optical quantum information systems, the complex photonic circuitry can operate at room temperature, but then cryogenic single-photon detectors are needed~\cite{Bradler:18,Tan:19}. The prospect of on-chip sources and single-photon detectors -- plus no fiber-to-chip coupling loss -- could potentially justify the overhead of cooling the photonic subsystem.

  In this work, we explore the coupling of W-center photoluminescence (PL) to silicon-on-insulator waveguides and microring resonators (MRRs).
  In prior work on silicon defect centers, waveguides were not used because PL collection has always occurred directly above the structure being pumped, i.e., normal to the surface of the sample. Several of these devices were on silicon-on-insulator of dimensions that could potentially support waveguides~\cite{Bao:08,Chong:11,Buckley:19arxiv}. Other studies have shown coupling of defect centers to suspended microdisk~\cite{Radulaski:15} and photonic crystal~\cite{Shakoor:13,Sumikura:14} cavities; however, resonator coupling of W~centers has not been shown.
  Reference~\cite{Kuznetsova:12} claimed W~center coupling to an undercut microdisk, but spectra were not recorded.
  W~center waveguide-coupling has been achieved before in an electrically-pumped device~\cite{Buckley:17}.
  Although this work aims to support the development of electrically-pumped sources, the setup for optical pumping and optical collection presents complementary advantages for research, further discussed below.
  Compared to Ref.~\cite{Buckley:17}, the waveguides are now single-mode, allowing study of a variety of photonic structures beyond a straight waveguide. Most notably, here we show coupling of W~center photoluminescence to modes of a MRR, which is coupled to a simple circuit of single-mode silicon waveguides. This result implies a potential to integrate W~centers in more complex waveguide circuits.

  To demonstrate coupling of PL to the silicon waveguide, we use an offset collection technique similar to techniques that have been used in III-V quantum dot research~\cite{Koseki:09,Englund:09,Coles:14}.
  When a structure is pumped, it emits upwards as well as into the waveguide. The upwards emitted component is blocked in an intermediate imaging plane. The waveguide-coupled light is routed to a normal-emitting grating coupler (GC) that is offset from the pump location. Since the GC emission is spatially offset, it avoids the block. Results indicate a qualitative difference in the spectra of upwards emitted PL and offset emitted (i.e., waveguide-coupled) PL; namely, the offset spectra exhibit comb features that are unambiguously characteristic of microring resonators (see Fig.~\ref{fig:mrr_data}).


\section{Layout and Fabrication}
  The substrate is a 76~mm wafer with a 220~nm silicon device layer on 2~\textmu m buried oxide. Implants were masked by 600~nm thick positive-tone resist patterned by electron beam lithography. The wafer was then implanted with $^{28}$Si$^+$ at a commercial facility.
  The silicon pattern is masked by an electron beam patterned oxide hard mask. The 220~nm silicon is fully etched to the buried oxide with a reactive ion etch based on SF$_6$ and C$_4$F$_4$. Finally, the wafer was encapsulated by a 1.5~\textmu m oxide. The wafer was diced into 1~cm $\times$ 1~cm die, which were then annealed. Implant and anneal conditions were chosen to yield the optimum W~center brightness as determined in Ref.~\cite{Buckley:19arxiv}.

  Waveguides were designed to be 350~nm wide such that they are single-mode at 1220~nm. SEMs of the waveguides show an actual width of 390~nm and unexpectedly high edge roughness. Cutback tests indicate a single-mode waveguide loss of 18~dB/cm, which was significantly higher than expected due to a glitch in the etch. The peak brightness of the W-center zero phonon line (ZPL) in an unpatterned SOI region is measured to give 4,400~$\pm$~220 counts/s with a 300~$\mu$W pump, which agrees with different cooldowns of different samples on the same setup. Microrings were designed with coupling gaps of 100~nm and radii of $r(\mu m)=$[2, 4, 8, 10, 20, 30].

  \begin{figure}[tb]
    \includegraphics[width=.9\linewidth]{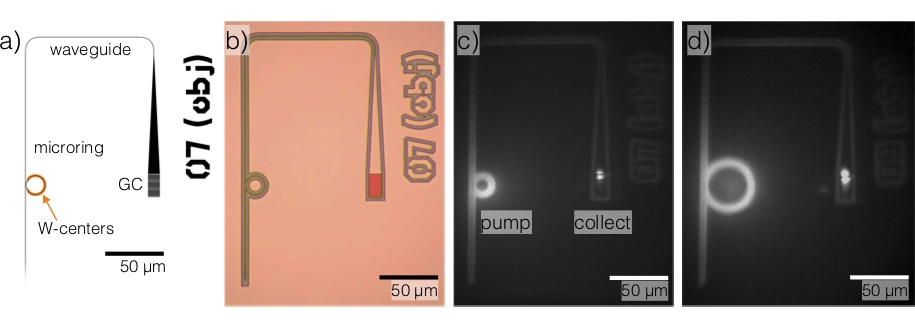}
    \caption{Different views of the circuit for offset collection from a microring resonator (MRR) with radius $r=$~8~\textmu m. a) Design layout. Silicon is black. The location of W-center implants is orange. GC: grating coupler. b) Visible microscope image of the fabricated circuit. c) Infrared image of the circuit when the MRR is optically pumped. Photoluminescence originates directly from the MRR (free-space coupled) and also from the offset grating coupler (waveguide-coupled). d) Infrared image of a different device with larger radius, $r=$~20~\textmu m.}
    \label{fig:circuit-view}
  \end{figure}

  We employ an offset collection technique that allows for observation of waveguide-coupled PL in silicon, which was introduced in our recent proceeding~\cite{Tait:20ofc}. The role of the silicon photonic circuit, shown in Fig.~\ref{fig:circuit-view}, is to spatially separate waveguide-coupled PL from free-space-coupled PL. The role of the optical apparatus is to isolate these two collection points. In the circuit, an MRR implanted with W~centers is evanescently coupled to a waveguide that routes waveguide-coupled PL to a grating coupler (GC). The GC is designed to emit normal to the surface and is located 100~\textmu m away from the MRR center. When the MRR is pumped, some component of the PL is emitted upward into free-space with a spectrum similar to that of unpatterned SOI. Another component of the PL couples to the MRR modes. This component eventually couples to the waveguide and is then routed to the GC. Since this GC is offset from the free-space emission source, the waveguide-coupled component is spatially discernible by the setup.

\section{Setup}
\begin{figure}[htb!]
    \includegraphics[width=.85\linewidth]{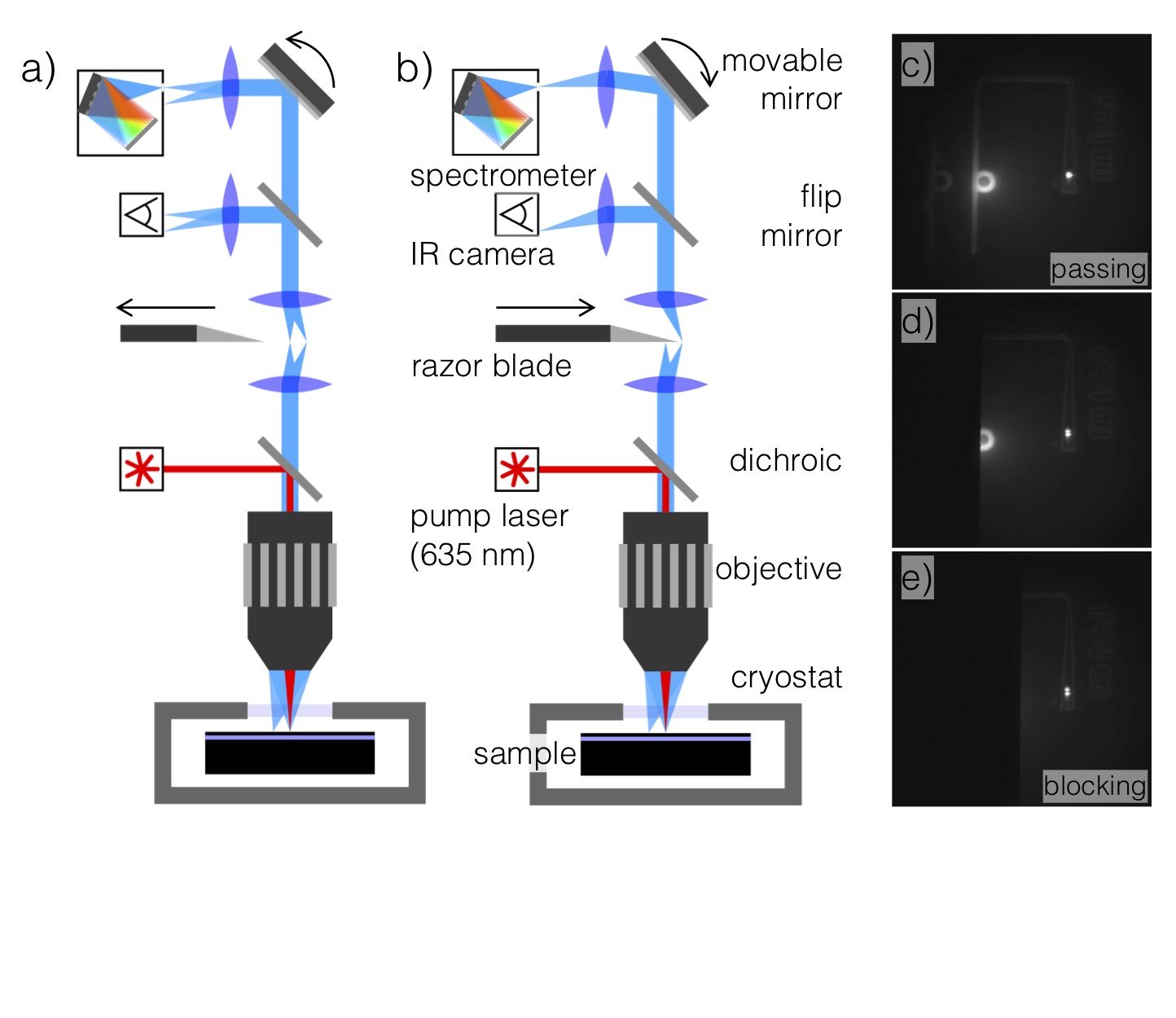}
    \caption{Simplified experimental diagram and visualization of the center PL blocking technique. a-b) Setup in non-blocking (a) and blocking (b) states. A pump laser (red) is focused onto a W-center device. Infrared (IR) PL is collected by the same objective. The returning PL transmits through a dichroic mirror, a razor blade assembly. Then the PL is split to an IR camera and spectrometer. When the center is blocked, the mirrors preceding the spectrometer are adjusted such that the offset component enters the slit of the spectrometer. c-e) Verification of the razor blade technique for isolating offset PL. c) Razor non-blocking corresponding to (a); d) razor halfway across MRR; e) razor fully blocking, corresponding to (b). Data is collected in the blocking state. Refer to the text for setup components that are not diagrammed.}
    \label{fig:setup}
  \end{figure}

  A simplified experimental setup is shown in Fig.~\ref{fig:setup}(a, b). The primary components are a pump, cryostat, objective, and infrared (IR) spectrometer. The 11~mW laser pump at 635~nm (red) is focused onto the device in the cryostat. Electrons and holes in the device recombine at W~centers and then emit IR photoluminescence (PL) (depicted as blue). This PL is directed to a spectrometer and InGaAs camera. PL is emitted from two locations: the device itself and the offset GC. When the free-space-coupled PL is imaged with a lens, the two emission sources are spatially separated. These distinct emission sources are seen in Fig.~\ref{fig:setup}(c).

  The objective performs both roles of focusing the pump to a spot and collecting the PL through a window in the cryostat. The long working distance objective has 10x magnification and a numerical aperture of 0.42. It is infinity corrected, meaning that the returned light is collimated and can traverse a long optical path before being re-imaged by a lens.
  The sample is fixed, and the objective is mounted on XYZ stages. Three mirrors (not shown) are mounted to the stages in a periscope configuration that maintains a consistent relationship between beams on the optical table and the objective, despite the movement of the objective.

  The procedure for discerning waveguide-coupled PL with this setup is as follows. Initially, the setup is aligned while collecting the free-space-coupled PL from the central pump location. Then, a razor blade in an image plane is moved to block the center of the field of view, seen in Fig.~\ref{fig:setup}(d). The blade is translated by a further 500~\textmu m corresponding to 50~\textmu m on chip. Fig.~\ref{fig:setup}(b) shows the beams and Fig.~\ref{fig:setup}(e) shows the IR image when the blade is in the blocking state.
  Once the center is blocked, mirrors before the spectrometer are adjusted until the offset signal is seen. Since only the offset PL makes it past the blade, we can be confident that this is the desired waveguide-coupled signal.

  Not shown in Fig.~\ref{fig:setup} are 1) mirrors in the XYZ alignment stage, 2) a pair of movable lenses after the pump allowing independent adjustment of the pump and PL focal planes, 3) a second mirror at each depicted mirror for independent beam angle/position alignment, 4) long-pass filters before IR camera and spectrometer to block the pump, 5) visible camera used for alignment, 6) tap and power meter for measuring pump power.

  The optical cryostat is a repurposed cryopump. To adapt the cryopump for the desired optical measurements, the top flange of the cryopump chamber was removed, and a machined brass post and brass sample holder were attached to the 15~K stage of the cold head. A radiation shield canister with a window was attached to the 80~K stage. Lastly, a new top flange with a window was put in place to seal the vacuum chamber. The cold head is driven by a closed-cycle helium compressor. A silicon diode thermometer is attached to the bottom of the sample holder, which reached a minimum temperature of 19.8~K.
  The compressor causes substantial vibration, meaning that optical measurements of small devices cannot be taken while it is running. We estimate a sample vibration displacement between 50 and 100~\textmu m. We address this vibration problem by turning the compressor off during fine alignment and measurement. While it is off, the temperature begins to increase, reaching 40~K in approximately 12 minutes. W-center brightness drops by 12\% between 28~K and 40~K and then drops off precipitously above 45~K~\cite{Davies:87,Buckley:19arxiv}. All measurements in this work were taken between 28~K and 40~K.

\section{Results}
  \begin{figure}[htb!]
    \includegraphics[width=.95\linewidth]{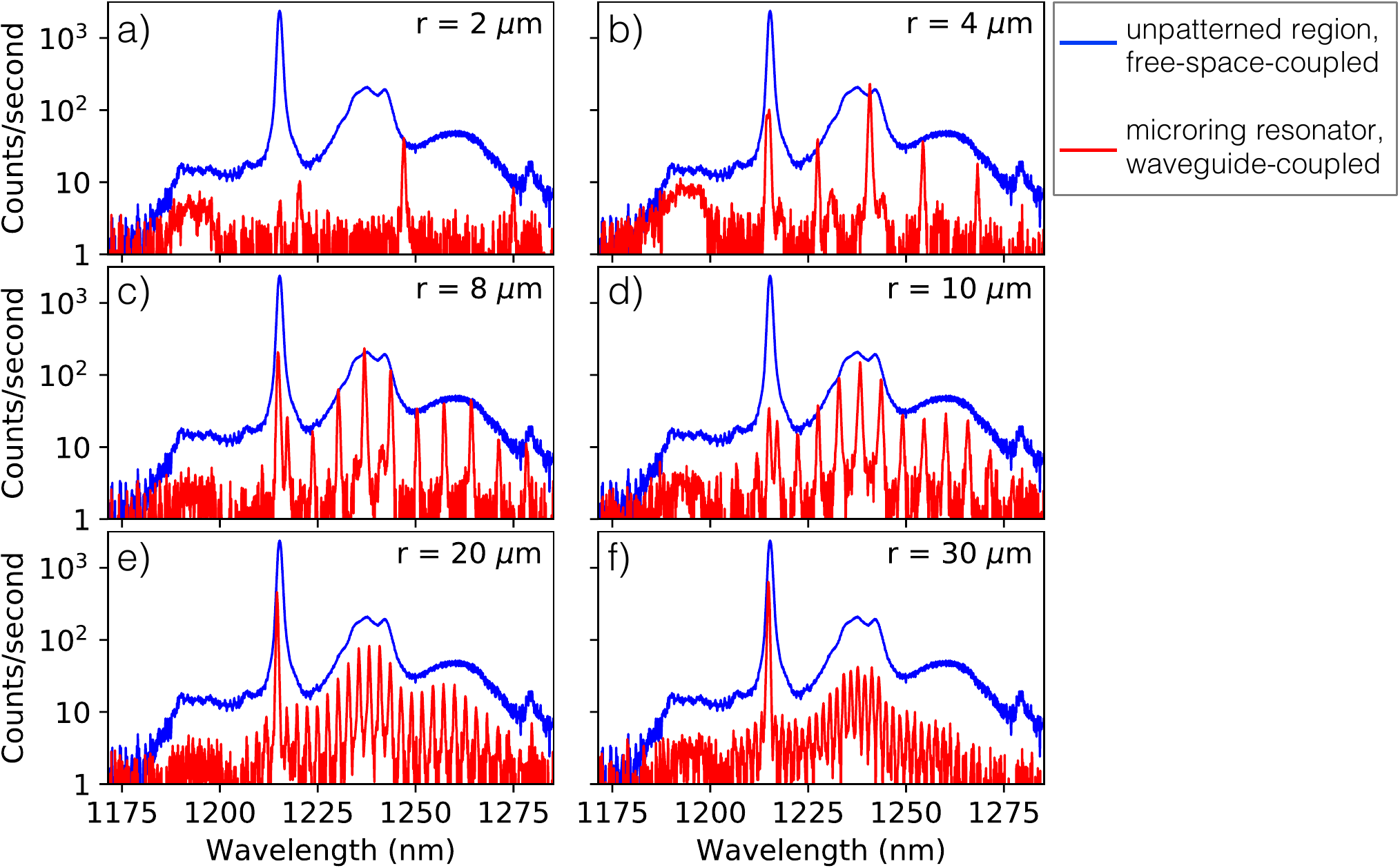}
    \caption{Comparison of PL from an unpatterned control region (blue curves) to PL from the GCs connected to MRRs of different radii (red curves). These curves correspond to free-space-coupled and waveguide-coupled PL. Multiple peaks at different wavelengths are indicative of multiple MRR resonances, which are evenly spaced. As MRR radius increases, the free spectral range decreases. The zero phonon line amplitude depends on where its wavelength falls relative to a resonator mode.}
    \label{fig:mrr_data}
  \end{figure}

  Waveguide-coupled spectra from MRRs with six different radii are shown in Fig.~\ref{fig:mrr_data}. These spectra (red) are compared to that of the control region (blue), which is an unpatterned, implanted SOI region. While W-center emission is brightest at the zero phonon line (ZPL) at 1218~nm, emission in the phonon sideband, from 1225~nm--1275~nm, is significant: in this measurement, 20~dB above the background. The sideband emission -- measurable with our setup -- allows us to spectrally resolve emission and transmission spectra over a broad, 50~nm wavelength range. It therefore can be used as a new diagnostic technique for a wide range of cryogenic silicon photonic devices.

  The collected PL is brighter at wavelengths that resonate with the microresonator and nearly undetectable off-resonance. This result indicates that only the PL emitted into the resonant modes couples to the bus waveguide.
  Preferential waveguide coupling does not necessarily mean preferential emission into the resonant modes, as in the sense of stimulated emission or Purcell enhancement, of which we did not observe conclusive evidence.
  The shape of the envelope of the peaks closely follows that of the W-center sideband from the control region. This indicates that the wavelength dependence of evanescent couplers and GC are relatively flat over the window. The data are not normalized, and it is coincidental that the amplitudes (dependent on multiple, competing, unmeasured factors, discussed below) also match.
  In Fig.~\ref{fig:mrr_data}(f), the peak amplitudes are significantly weaker because the MRR diameter is much larger than the focused pump spot. The pump is defocused so that more of it reaches the MRR waveguide; however, most of the pump power still passes through the unimplanted center.

  The ZPL is visible in every sample, but its amplitude varies considerably. Its amplitude depends on how close a MRR resonance falls to 1218~nm. As a result of resonance sensitivity to fabrication variation, it is nearly impossible to control the absolute resonant wavelengths~\cite{Lu:17}. On the other hand, the relative spacing between resonances, i.e., free spectral range (FSR), is precisely controlled by MRR radius according to the formula
  \beq
    FSR = \frac{\lambda^2}{2\pi r \cdot n_g},
  \eeq
  where $n_g$ is the group index. There is close agreement between this model and the data. The best fit of the measured FSRs from Fig.~\ref{fig:mrr_data} yields $n_g = 4.51$ with a fit residual of $R^2=0.991$.
  Previous measurements of similar MRRs, taken using a transmission spectrum analyzer and fiber alignment setup, yielded a group index of 4.33$\pm$0.03; however, that measurement was at room temperature and without W~centers.

\begin{figure}[tb]
    \includegraphics[width=.85\linewidth]{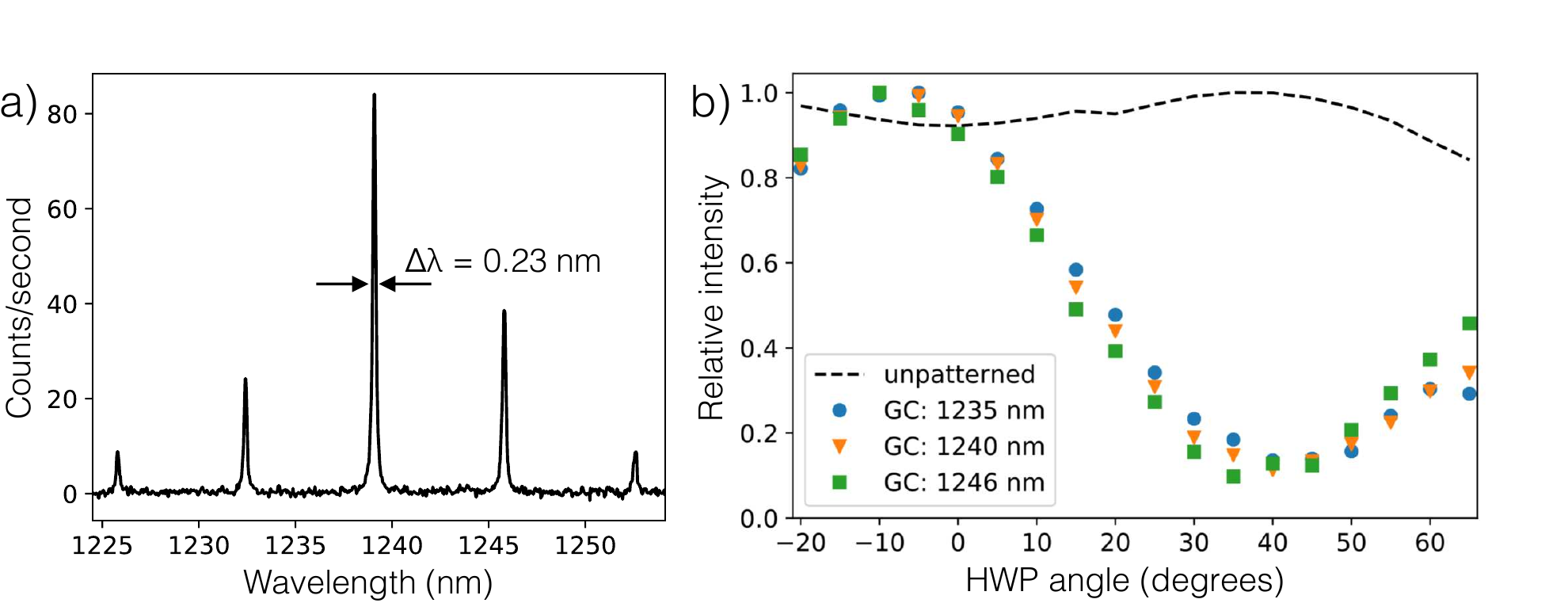}
    \caption{a) High resolution spectrum of the sideband of the MRR with $r=$~8~\textmu m (Fig.~\ref{fig:mrr_data}c). The Q-factor of the central peak is at least 5,300. b) Polarization dependence of light collected from the grating coupler (GC) compared to unpatterned control region. The GC curves are derived from 3 peaks of the $r=$~10~\textmu m microring (Fig.~\ref{fig:mrr_data}d). Each curve is normalized to its maximum, and the values collected from the GC are first normalized by the corresponding unpatterned value. Note that (a) and (b) are different devices.}
    \label{fig:highres_and_pol}
  \end{figure}

  Figure~\ref{fig:highres_and_pol}a shows a high resolution waveguide-coupled spectrum of the $r=$~8~\textmu m MRR. A spectrometer grating with 1,200~grooves/mm is used for this measurement compared to the grating used for Fig.~\ref{fig:mrr_data} with 300~grooves/mm. The full-width half-maximum (FWHM) of the center peak is measured to be $\Delta\lambda=$~0.23~nm; however, this is close to the theoretical resolution limit of the detector array. Since any kind of misalignment or defocus can affect the actual resolution limit, it is possible that the FWHM is overestimated. The quality factor of this feature is at least 5,300. The intrinsic Q of this MRR is at most 55,000 due to waveguide propagation loss, which is measured to be 18~dB/cm. The extrinsic Q is determined by the combined round trip propagation loss plus round-trip waveguide coupling coefficient, which is simulated to be $\kappa^2=$~0.121. The extrinsic Q is therefore expected to be at most 7,700. Measurement and simulation place relatively tight bounds on the real Q between 5,300--7,700. In~\cite{Tait:20ofc}, measurements of a microdisk on the same setup and sample found resonances with Q-factors up to 7,160. The higher Q can be explained because microdisk modes couple more weakly to the bus waveguide and interact less with any etched sidewalls.

  Figure~\ref{fig:highres_and_pol}b shows the PL polarization. To measure polarization, a half wave plate and polarizer are placed before the razor blade assembly. The allowed polarization angle is twice the half wave plate angle. Values labeled ``GC'' refer to the amplitudes of different peaks in the waveguide-coupled sideband of the $r=$~10~\textmu m MRR (Fig.~\ref{fig:mrr_data}c). Unpatterned values refer to the ZPL amplitude of an unpatterned control region. The degree of polarization of the GC emission is substantially stronger (88\% depth) than that of the control (7.8\% depth), which is corroborative because the GC is designed to couple only TE light.
  A half wave plate angle of 0 means that S-polarization (E-field perpendicular to the dichroic surface is nonzero) is detected. The dichroic mirror causes a polarization dependence of 3.3\% (characterized by manufacturer at 45$^\circ$ incidence at 1218~nm).
  The sample is mounted roughly parallel to the setup such that TE polarized light in its waveguides ends up as S-polarized in the setup. The maximum of GC traces actually occurs at $-10^\circ\pm5^\circ$. Most likely, this means that the sample was rotated by this much away from parallel.

\section{Discussion}
  Previous work on waveguide-coupled W-center emission used electrical pumping and on-chip detection~\cite{Buckley:17} whereas this work uses free-space optical pumping and collection techniques. The integrated and free-space approaches present complementary advantages for research.
  As a consequence of optical pumping, a whole array of devices can be probed in a single cooldown. Photoluminescent properties can be studied independent of electrical properties such as ohmic heating. Finally, fabrication is two-step (Si rib etch, Si$^+$~implant) instead of seven-step needed for electrical injection (Si rib etch, Si partial etch, Si$^+$~implant, heavy/light boron, heavy/light phosphorus)~\cite{Buckley:17}.
  In addition to these optical pumping advantages, optical collection provides other complementary advantages. Superconducting detectors are not required, meaning that operating temperature can be 40~K instead of 1~K. Free-space collection allows use of commercial detection instruments (e.g., spectrometers, high-speed diodes, cameras) without a need for fibers in the cryostat. The presented techniques could thus play a central role in the characterization of cryogenic silicon photonic platforms. For example, we have observed the FSRs of various MRRs and precisely determined their group velocity by examining the sideband spectrum, while any sort of on-chip detector could only respond to the integrated luminescence over the band.

  In several cases, it would be desirable to make a resonance align with the bright ZPL. This alignment presents a challenge because both the resonances and the ZPL are narrow, and the resonant wavelengths are highly sensitive to fabrication variability. Resonance alignment is required for lasing. We did not observe conclusive evidence of lasing in the devices in this study primarily because of the resonance alignment challenge.
  This challenge can be addressed by either making cavities with small FSRs or through refractive index tuning.

  Cavities designed with FSRs smaller than the ZPL linewidth could ensure that one resonance aligns with the ZPL.
  In this work, the $r=$~30~\textmu m MRR has an FSR that meets the condition. The problem with this device is that it is much larger than the pump spot size, so nearly all of the pump power passes through the center of the ring, missing the implanted waveguide. We fabricated a $r=$~40~\textmu m device, but it was not measurable for this reason. Next steps could include modifying the ring shape or increasing pump power further.
  In general, the approach of using small-FSR cavities has the issue that they must be physically large, meaning that they have undesirably large footprint and capacitance. Additionally, without a means of tuning, parasitic index changes (e.g., from electrical pumping) cannot be countered.

  Refractive index tuning can be used for static and dynamic resonance alignment. It is typically implemented with heaters at room temperature. Below 4~K, however, thermooptic tuning is not viable because $dn/dT$ approaches zero~\cite{Komma:12}. There are other candidate approaches to refractive index tuning at cryogenic temperature to explore in further work.
  Index tuning can be accomplished with a PN junction, although this effect is weak ($\sim$100~pm total range in~\cite{Gehl:17}).
  It may also be possible to realize a sufficient index change by locally heating the W-center/cavity device between 4~K and 40~K.
  Yet another approach is electromechanical tuning. Simulations of a photonic crystal from~\cite{Yan:18} indicate a potential tuning range of more than 10~nm; this device has not yet been experimentally demonstrated.
  Static adjustment of MRR resonances could be realized with defect-mediated or other postfabrication trimming~\cite{Ackert:11,Alipour:15,Chen:18}, although these techniques are not capable of dynamic adjustment.

  Comparing the shape of the control spectrum to the envelope of GC peaks in Fig.~\ref{fig:mrr_data}, it appears that the GC response is mostly flat over the phonon sideband of interest.
  This data could be slightly improved by characterizing the GC response shape.
  As seen in Fig.~\ref{fig:circuit-view}(c, d), emission appears to originate entirely from the front of the GC, and sometimes the emission pattern is bifurcated into two lobes, which indicates that this GC is not optimal.
  The current setup would be ideal for GC characterization and optimization in further work: using an array of identical sources coupled to varying GCs, the GC that is optimal for objective-coupled normal emission at 1218~nm at 20~K could be found in a single cooldown.

  This work presents several directions for further study.
  Relative emission into the TM polarization could be characterized using different GC designs for TM coupling. Source lifetime is a critical parameter for superconducting optoelectronic neural networks~\cite{Shainline:19}. Lifetime engineering could perhaps be accomplished by modifying the optical cavity density of states, as in Ref.~\cite{Sumikura:14}. Engineering W-center density to be on the order of $\lambda^{-3}$ (one emitter per wavelength volume) could be a route to silicon single-photon sources; however, currently, the density of W~centers has not been measured, let alone as a function of implant parameters.

  Of particular interest would be investigations into a W-center laser. A W-center laser would have a variety of applications for on-chip optical measurement and nonlinear optics, for example, photon pair generation through degenerate four wave mixing~\cite{Gentry:15,Savanier:16}. Stimulated emission has been observed from G-center defects in nanopatterned silicon~\cite{Cloutier:05}; however, it remains to be shown whether population inversion -- let alone amplification -- can be achieved in a waveguide that is implanted with W~centers.
  Another interesting prospect would be using the broad sideband for wavelength-division multiplexed (WDM) sources. The sideband is 30 times less bright than the ZPL, but this may still be efficient enough for neuromorphic silicon photonic architectures, which depend critically on photonic routing networks~\cite{Shainline:17,Tait:14}.

  In conclusion, we demonstrated microresonator- and waveguide-coupled emission from silicon W~centers. Light at resonant wavelengths that is emitted into the modes of each resonator is preferentially coupled to an adjacent waveguide. By examining spectral features in the 50~nm wide phonon sideband, we were able to determine each device's FSR, which closely agree with the model of microring FSR vs. radius. The Q-factor of one device is determined to be at least 5,300, and the polarization is measured to be 88\% as a result of GC polarization dependence. W~centers are promising candidates as straightforward, 300~mm wafer-scale, waveguide-integrated silicon light sources. These results and methods open numerous directions for further research of fundamental defect center properties, all-silicon active devices, and large-scale photonic information processing.

  \

  We thank R.~Boutelle and K.~Silverman for helpful discussions on the experimental setup. We thank S.-P.~Yu and N.~Sanford for editorial input.

  \


\vfill
\bibliographystyle{iopart-num}
\bibliography{iop-emission}

\end{document}